\documentclass[12pt]{article}

\usepackage[english]{babel}
\usepackage[utf8]{inputenc}
\usepackage{amsmath}
\usepackage{empheq}
\usepackage{amssymb}
\usepackage{graphicx} 
\usepackage[font=small,labelfont=bf, width=0.95\textwidth]{caption}

\usepackage{mdframed} 
\usepackage{framed} 
\usepackage[most]{tcolorbox} 
\tcbuselibrary{breakable}

\usepackage{booktabs} 

\usepackage{natbib}

\usepackage{lineno} 

\title{Time-structured models of population growth in fluctuating environments}
\author{ \normalsize{Pradeep Pillai$^{1,\dagger}$ and Tarik C. Gouhier$^{1}$}\\
\normalsize{$^{1}$Marine Science Center, Northeastern University,}\\
\normalsize{430 Nahant Rd, Nahant, MA 01908}\\
\normalsize{$^{\dagger}$Corresponding author: pradeep.research@gmail.com} }

\date{}

\begin{document}

{\centering

\textbf{\Large{Time-structured models of population growth in fluctuating environments}} \\
\bigskip
Pradeep Pillai$^{1,\dagger}$ , Tarik C. Gouhier$^{1}$ \\
\textit{
$^{1}$Marine Science Center, Northeastern University,\\
430 Nahant Rd, Nahant, MA 01908\\
$^{\dagger}$To whom correspondence should be addressed. E-mail: pradeep.research@gmail.com \\
}
}

\begin{abstract} \noindent \begin{enumerate} 
  \item Although environmental variability is expected to play a more prominent role under climate change, current demographic models that ignore the differential environmental histories of cohorts across generations are unlikely to accurately predict population dynamics and growth. The use of these approaches, which we collectively refer to as \textit{non time-structured models} or nTSMs, will instead yield error-prone estimates by giving rise to a form of ecological memory loss due to their inability to account for the historical effects of past environmental exposure on subsequent growth rates.  
  \item To address this important issue, we introduce a new class of \textit{time-structured models} or TSMs that accurately depict growth under variable environments by splitting seemingly homogeneous populations into distinct demographic cohorts based on their past exposure to environmental fluctuations. By accounting for this cryptic population structure, TSMs accurately simulate the historical effects of environmental variability, even when individuals exhibit different degrees of phenotypic plasticity.
  \item Here, we provide a conceptual framework, the mathematical tools needed to simulate any TSM, and a closed form solution for simple exponential growth. We then show that traditional nTSMs yield large errors compared to TSMs when estimating population dynamics under fluctuating temperatures. Overall, TSMs represent a critical tool for predicting population growth in a variable world.
\end{enumerate}

\end{abstract}


\section{Introduction}

Environmental variation is expected to have strong effects on the dynamics of marine and terrestrial ecosystems over the course of the 21$^\text{st}$ century \citep{walther_ea_02,walther_10, harley_ea_06}. For instance, the mean, variance and autocorrelation of temperature fluctuations are all projected to increase over time \citep{dicecco_gouhier_18,duffy_ea_22,lenton_ea_17} and ultimately lead to reduced organismal performance \citep{vasseur_ea_14} and higher extinction risk \citep{duffy_ea_22}. It is thus critical that we develop empirical and theoretical approaches that can properly estimate the effects of temporal environmental variation on the size, growth and persistence of ecological populations. Unfortunately, studies of taxa ranging from tadpoles and killifish to insects suggest that scaling-up from individuals to populations may be difficult because historical effects can make it impossible to infer organismal performance such as growth under temporal environmental variation from their performance under the relevant constant conditions \citep{kingsolver_ea_15,niehaus_ea_12,denny_19,angilletta_09,khelifa_ea_19,kingsolver_woods_16}.

Although more likely to occur in endotherms due to their internal buffering mechanisms, historical effects can also arise in ectotherms because even the simplest organisms typically do not respond instantaneously to environmental changes \citep{kingsolver_ea_15,niehaus_ea_12}. Hence, in most cases, the sequence of environmental fluctuations an organism experiences in the past will influence its response to current and future conditions. These historical effects are critically important because they can result in repeated exposures to sublethal environmental conditions either increasing or decreasing the likelihood that an organism will survive potentially lethal future conditions \citep{angilletta_09}. For instance, exposure to elevated but sublethal temperatures can prime corals to survive subsequent temperatures that would ordinarily elicit bleaching and mortality \citep{ainsworth_ea_16}. 

Recent theoretical and empirical work has begun to scale-up from individuals by developing mathematical models and experimental protocols to determine the impact of historical effects that arise under temporal environmental variation as well as the mediating role of phenotypic plasticity, whereby acclimation can allow organisms to offset performance losses incurred under suboptimal conditions \citep{kremer.fey.ea_18,bernhardt_ea_18,fey_ea_21,gill_ea_22}. Although they differ in some respects, these approaches, which we refer to collectively as non time-structured models or nTSMs, all share a common and critical \emph{implicit} assumption: populations are considered homogeneous in that all individuals are presumed to be functionally identical despite having experienced different sequences of environmental conditions. Hence, by assuming that organisms born in and acclimated to different environments all respond in perfect unison, these approaches ignore the emergence of a kind of cryptic population structure that can lead to large errors when predicting population growth under fluctuating conditions \citep{gouhier.pillai_19}. Effectively, any approach that treats populations undergoing temporally variable environmental conditions as homogeneous will introduce a form of ecological memory loss that is likely to lead to major mistakes \citep{padisak_92}.

In order to avoid the emergence of such ecological memory loss, we introduce a novel framework called time-structured models or TSMs, which can be used to study population growth in fluctuating environments. This framework accounts for the way that population growth varies as the environment cycles through distinct discrete states when organisms exhibit some degree of acclimation. Specifically, since a population's growth rate is mediated by its constituent organisms who are born at different times and thus experience distinct sequences of environmental conditions, the population will be temporally structured into a series of homogeneous cohorts each characterized by a unique environmental history and acclimation response. This means that any mathematical description of total population growth needs to account for the distinct historical effects and acclimation responses associated with each cohort as a result of their exposure to different sequences of environmental conditions.

Here, we provide both a conceptual and a mathematical framework based on ordinary differential equation (ODE) systems that can track a temporally structured population. The differential equations of the TSM framework are amenable to being solved explicitly when population growth is sufficiently exponential in nature \citep{kremer.fey.ea_18,fey_ea_21}. Below, we present the overarching TSM framework and then derive explicit closed form solutions when population growth is exponential. Then, in order to demonstrate the robustness of TSMs, we consider how the expressions for the general solution transforms under specific scenarios represented by two limiting cases where (i) the entire population is homogeneous and (ii) there is no phenotypic plasticity, in that individuals are unable to acclimate to new environmental conditions. In the homogeneous population case, all the individuals in every cohort born under different environmental regimes are indistinguishable from each other, and thus all acclimate and respond to environmental changes in unison (i.e., the population is memory-less with regard to past environmental variation). This limiting case corresponds to the implicit assumption made when modeling population growth for experimental studies \citep{kremer.fey.ea_18}. The second scenario without phenotypic plasticity provides a suitable null model to study how short-term temperature or environmental adaptability can modify growth under environmental variation. Finally, we close with a case study showing the magnitude of the errors that can arise when using nTSMs to estimate population growth and size under fluctuating temperatures.

Overall, the TSM framework offers a novel method for studying population growth in laboratory and field populations. The general framework can be instantiated into multiple specific forms to create a family of TSMs, each tailored to the specific needs or assumptions of a given experimental system. The closed form solutions for the ordinary differential equation system itself provides a powerful and timely extension to classical population biology models for accurately predicting the effects of environmental change.

\section{Population growth in time-structured models}
The general TSM framework accounts for the effects that changes in some environmental condition such as ambient temperature will have on population growth. In TSMs, the total heterogenous population is split into distinct homogeneous subpopulations or cohorts, each of which is composed of individuals born and acclimated during a particular time period characterized by a discrete environmental state. If cohorts are defined as homogeneous subpopulations consisting of individuals created under the same environmental regimes, then it would be reasonable to expect that distinct demographic cohorts within a population will be acclimated to different temperature conditions, and as a result, will likely exhibit different plastic responses to temperature changes (i.e., rates of temperature acclimation) despite being part of the same overall population.

\begin{figure}
  \centering
    \includegraphics[width=0.55\textheight]{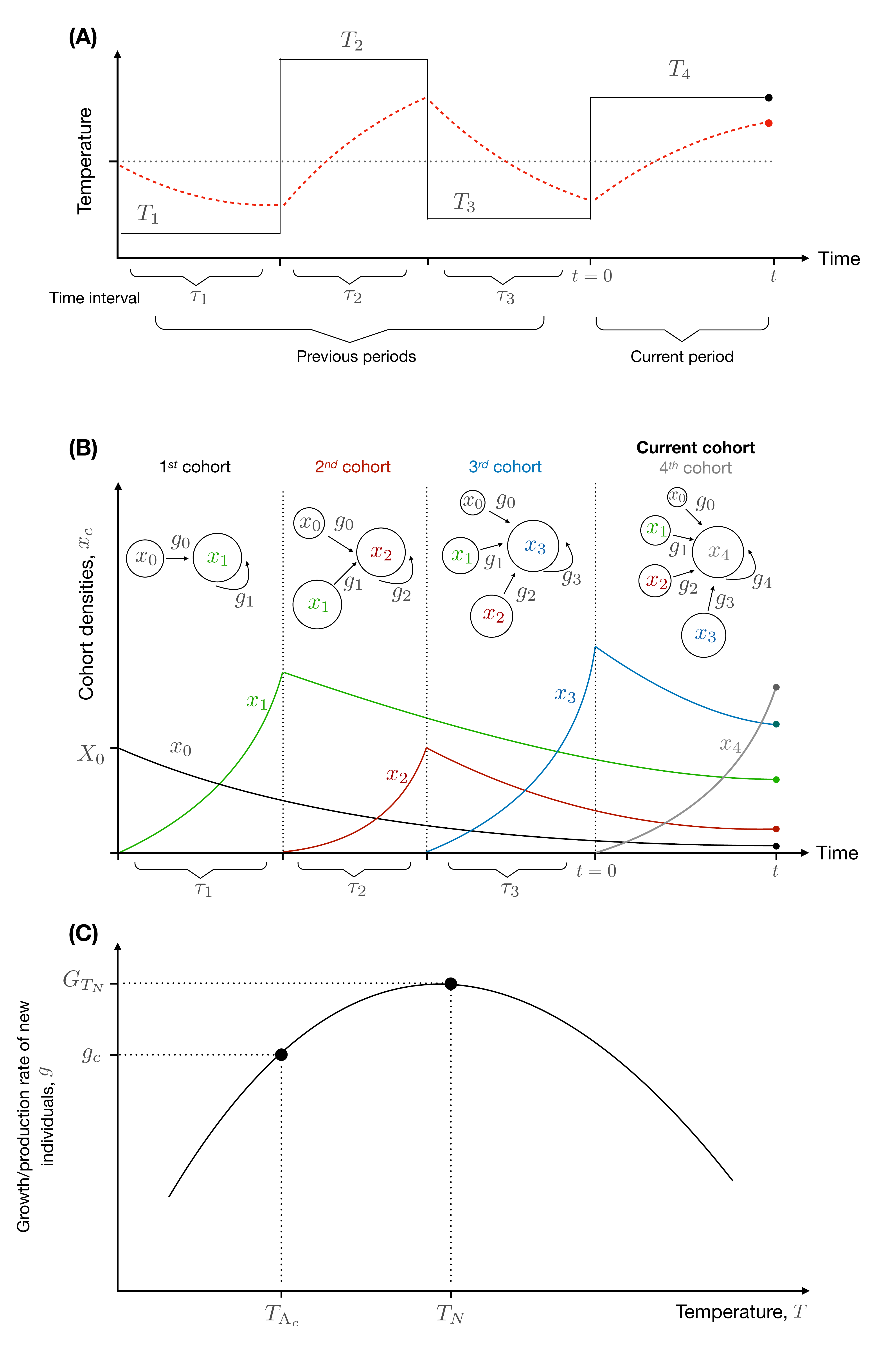}
    \caption{\textbf{Time-structured models track cohort densities}. (A) Temperature acclimation of initial cohort across multiple temperature states (red dashed line) compared to initial acclimation temperature (black dotted line). (B) Time series of population cohort densities. (C) An example of a cohort's reproduction rate, $g_c$, as a function of its acclimation temperature, $T_{A_c}$.}
    \label{Fig:fig1}
\end{figure}

Because cohorts in the same overall population potentially behave differently under distinct temperature regimes, if we wish to measure total population density, we would need to track the densities of each demographic cohort at any given time during the population growth phase. When mathematically formulating our population growth model as an ordinary differential equation system, we measure total population density, $x(t)$, at time $t$, and define time $t$ as being the time the system has been in the \textit{current} temperature state, $T_{\mathrm{Current}}$. The time at the beginning of the current environmental/temperature state is thus $t=0$, and is the point in time that determines the initial values for all the density variables in the ODE system. For all of the \textit{past} temperature periods, the duration of time spent in each of the previous temperature states ($T_i$) is given by $\tau_i$ (Fig. \ref{Fig:fig1}A). 

Tracking the densities of each cohort, both past and present, entails recording birth and death processes while in different temperature states. If demographic cohorts are simply defined by the time period or temperature state during which individuals are born, then the individuals that are reproducing and thus contributing to the \textit{current cohort} generation will not only be individuals belonging to the current cohort, but also any reproductive individuals belonging to all previous cohorts. Once a time period or temperature state has passed, the cohort associated with that state can no longer grow in density, but begins to decay as individuals in that cohort die off (Fig. \ref{Fig:fig1}B). 

Next, we will provide an overview of how to formulate the demographic parameters that determine the birth and death processes in a time-structured model. Afterwards, we show how to both set-up and solve the relevant ordinary differential equation systems for describing population growth under varying environmental states.

\subsection{General growth model}
A growing population that has experienced a temporal sequence of $N$ successive and distinct environmental/temperature states over a given time span can be split into $N$ time periods (with the $N^{\text{th}}$ time period/state representing the current one). Let us now assume that at any given point in time, the total population of individuals can be subdivided into demographic cohorts based solely on what time period or environmental/temperature state the individual was originally born in. Thus, with $N$ time periods or temperature states, we will have a total of $N+1$ cohorts, including the zero cohort representing the initial subpopulation at the start of the time series. The change in total population density at a time $t$ in the current period ($N$) can then be represented by the differential equation
\begin{align}
  \frac{dx_{\mathrm{total}}(t)}{dt} &= \frac{dx_{0}(t)}{dt} +\frac{dx_{1}(t)}{dt} + \cdots +  \frac{dx_{N}(t)}{dt}.
\label{eq:total_dynamics_2cohort}
\end{align}

For for any cohort $c$, the birth, $B$, and death, $D$, rates determining the change in $x_c(t)$ in a simple logistic growth model can be represented by
\begin{align}
  B_c &= g_c \, x_c - x_c\sum_{i=0}^{N} \xi_{c,i} \, x_i,  \notag \\
  D_c &= \delta_c\, x_c +  x_c \sum_{i=0}^{N} \zeta_{c,i} \, x_i, \notag 
\end{align}
where $g_c$ and $\delta_c$ are the intrinsic (density-independent) growth and mortality rates, while $\xi$ and $\zeta$ are the slopes of the curves indicating the density effects of all cohorts on the birth and death rates of the $c^{\text{th}}$ cohort. 

The dynamic equations for a population with $N+1$ cohorts in a density-dependent model (with $r=g-\delta$ and $\alpha_{c,i}= \xi_{c,i}+\zeta_{c,i}$) are
\begin{align} \label{eq:TSM_ODEnonlinear}
  \frac{ d x_0}{dt} \,&=\, -x_0 (\delta_0 + \sum_{i=0}^{N} \zeta_{0,i} \, x_i), \notag \\
  \frac{ d x_1}{dt} \,&=\, -x_1 (\delta_1 + \sum_{i=0}^{N} \zeta_{1,i} \, x_i),   \notag \\
  &\,\, \vdots \notag \\
  \frac{ d x_N}{dt} \,&=\, x_N \left(r_N  -\sum_{j=0}^{N} \alpha_{c,j} \, x_j \right) + \sum_{i=0}^{N-1}  x_i  \left(g_i  - \sum_{j=0}^{N} \alpha_{c,j} \, x_j  \right) .
\end{align}
In the above ODE system, the rate equations for all the previous cohorts, $\dot{x}_0$ to $\dot{x}_{N-1}$, simulate decaying densities, which is to be expected given that each cohort's reproductive output $g_c$ only contributes to the current $N^{\text{th}}$ cohort. Since the reproductive output from all cohorts only enter or contribute to the density of the current or $N^{\text{th}}$ cohort, it is only in the $N^{\text{th}}$ cohort equation that we see density being affected by reproduction. In the equation for the $N^{\text{th}}$ cohort (i.e., $\dot{x}_{N}$), the first term accounts for the \textit{net} contribution of new progeny produced by the current cohort, while the second term sums the contributions made from all other past cohorts. 

Eq. \eqref{eq:TSM_ODEnonlinear} is a non-linear density-dependent time-structured system. We can simplify this model system to get the density-independent version, as given by
\begin{align} \label{eq:TSM_ODE}
  \frac{ d x_0}{dt} \,&=\, -x_0 \, \delta_0,   \notag \\
  \frac{ d x_1}{dt} \,&=\, -x_1 \, \delta_1,   \notag \\
  &\,\, \vdots \notag \\
  \frac{ d x_N}{dt} \,&=\, x_N \, (g_N - \delta_N)  + \sum_{i=0}^{N-1}  x_i  \, g_i  .
\end{align}
This is the cohort version of exponential population growth (i.e., non asymptotic or non density-dependent growth), where all mortality $\delta_i$  is assumed constant, a situation which may occur when there is constant dilution or flushing from the system, as in a chemostat, or when densities are sufficiently far from equilibrium, as is often seen in large natural systems.

\subsection{Temperature-dependent parameters}
Let us assume there have been $N$ distinct temperature phases in a time series, and consequently at present the total population is made up of $N+1$ demographic cohorts (cohorts range from the $0^{\text{th}}$ to the $N^{\text{th}}$). The  temperature during the current phase being considered is $T_N$. We take the rate at which an individual originally born in one of the previous $c$ cohorts (where $c <N$) contributes to the current population, $g_c$, at time $t$ as
\begin{align} \label{eq:g_c}
  g_c(t) = G_{T_N} - b_N\Big(T_N - T_{A_c}(t)\Big)^2,
\end{align}
where $b_N$ is a parameter constant (which may be specific to the current temperature state, hence the subscript $N$), and $G_{T_N}$ is the maximum or optimum growth rate possible if individuals from the $c^{\text{th}}$ generation were perfectly acclimated to the current temperature $T_N$. This initial value of $g_c(t)$ at time $t=0$ (the beginning of the current period) will be determined by how the $c^\text{th}$ cohort was acclimated to the past temperature phases (for example, see Fig. \ref{Fig:fig1}A for how one cohort changes in its temperature acclimation across different temperature states).

The temperature $T_{A_c}(t)$ will be used to give the temperature that individuals in the $c^\text{th}$ cohort are currently acclimated to at time $t$. If we allow this acclimated temperature to change over time, such that individuals exhibit phenotypic plasticity, then this acclimated temperature will be a function of $t$, $T_{A_k}(t)$. The simplest way to model this phenotypic plasticity is for the acclimated temperature to change based on the difference between the current ambient temperature and the temperature individuals are actually acclimated to at any given time $t$:
\begin{align} \label{eq:dT_dt}
  \frac{d\,T_{A_c}}{dt} = v\Big(T_N - T_{A_c}(t)\Big),
\end{align}
for rate or velocity of change $v$. The solution to this  differential equation is
\begin{align} \label{eq:T_A}
  T_{A_c} &= T_N - \Big(T_N - T^{\,\circ}_{{A}_c}\Big)e^{-vt}, \notag \\
  &= T_N - \Delta_{_{c,N}} \, e^{-vt}.
\end{align}
The parameter constant $T^{\,\circ}_{{A}_c}$ is the \textit{initial} acclimated temperature of the $c^\text{th}$ cohort at the beginning of the current temperature phase (i.e., when $t=0$). Thus $\Delta_{c,N}$ is the parameter constant giving the \textit{initial difference} between the $N^\text{th}$ or current ambient temperature, $T_N$, and the temperature that the $c^{th}$ cohort is actually acclimated to at the \textit{start} of the current $N^\text{th}$ period, $T^{\,\circ}_{{A}_c}$, such that $\Delta_{c,N} = T_N - T^{\,\circ}_{{A}_c}$. Substituting Eq. \eqref{eq:T_A} back into Eq. \eqref{eq:g_c} gives us the rate at which members of a previous $c$ cohort contributes new individuals to the current cohort, 
\begin{align} \label{eq:g_c_explicit}
  g_c(t) = G_{T_N} - b_{_N}\,\Delta_{{c,N}}^2 \,\, e^{-2vt}.
\end{align}

\subsection{Simple example using a two-cohort model}
We wish now to find the explicit solutions to the differential equations describing a system exhibiting time-structured exponential population growth. To demonstrate the method for solving the differential equation system for the time-structured model, we will start with a simple two cohort population (i.e., with the zeroth and first cohort only),  and with only one temperature phase ($N=1$), and then extend this approach to a system with an arbitrary number ($N>1$) temperature states. For convenience, from here on in when deriving the explicit solutions to the ODE system, we will assume the loss rate $\delta$ is identical across cohorts (as in a large, constantly flushed chemostat). The same approach for solving the ODE system that we demonstrate below holds when this assumption is relaxed.

We will consider here a single temperature state, $T_1$, at the beginning of the time series that is different from whatever value occurred prior to the time series, and where the population density at the start of the time series at $t=0$ will be given by the parameter constant $X_0$. At any given time $t$ during the first temperature state, $T_1$, there will be two cohorts -- the zeroth cohort (with density $x_0(t)$), representing the subpopulation of individuals present from the start of the temperature shift or beginning of the time series at $t=0$, and the first cohort of individuals (density $x_1(t)$) that were \textit{created} during the first temperature state $T_1$. The densities of the zero and first cohort at time $t=0$ (beginning of the current temperature cycle) provide the initial conditions for solving the population density at time $t$. Since the first cohort represents individuals that were only created during the first temperature state, then the initial condition associated with this cohort will be $x_1(0) = 0$, while the other initial condition is given by $x_0(0) = X_0$. Thus, at time $t$ in the current temperature/environmental cycle we have an Initial Value Problem (IVP) given by
\begin{align} \label{eq:IVP_2cohort}
  \frac{ d x_0}{dt} \,&=\, - \delta_1 x_0(t),   \\
  \frac{ d x_1}{dt} \,&=\,  g_0(t;T_1) \cdot x_1(t) \,\, + \,\,  \left( g_1(T_1) - \delta_2 \right) \cdot x_1(t), 
\end{align}
for initial conditions $x_1(t=0) =0$, and $x_0(t=0) = X_0$. Here $g_0(t)$ is the growth rate of the initial cohort which was originally acclimated to a temperature $T_{A_0}$ prior to the current temperature $T_1$, and thus may potentially change over time as the individuals in this cohort attempt to acclimate to the current temperature regime $T_1$. Meanwhile $g_1$ is the growth rate of the first cohort of individuals born during the current time period, and are thus already acclimated to the current temperature, $T_1$.

\paragraph*{Density of the zero cohort}
Since the first equation in the ODE system above (Eq. \eqref{eq:IVP_2cohort}) has a clear solution, we can obtain an explicit expression for the density of the zero cohort, $x_0(t)$, after time $t$ in the temperature state $T_1$, 
\begin{align}
  x_0(t) =  X_0 \, e^{-\delta_0 t  }. 
\end{align}

\paragraph*{First temperature state}
Since the explicit solution for the density of the zero cohort is $x_0(t) = X_0 e^{-\delta t}$, this allows us rewrite the IVP above in Eq \eqref{eq:IVP_2cohort} as 
\begin{gather} \label{eq:simple_eg_dx1_dt}
  \frac{ d x_1}{dt} \,=\,  X_0 \, e^{-\delta_0 t}\cdot g_0(t; T_1) \,\, + \,\,  \left( g_1(T_1) - \delta_1 \right) \cdot x_1(t),   \\
  \text{initial condition:} \, \quad x_1(0) = 0. 
\end{gather}
This is a first order linear differential equation, which when put in the standard form of $\frac{dx}{dt}+P(t)\,x(t) = Q(t)$ yields
\begin{align}
  \frac{ d x_1}{dt}  \,\, - \,\,  \left( g_1 - \delta_1 \right) \cdot x_1(t)  \,\, = \,\,   X_0 \, e^{-\delta t} g_0(t)
\end{align}
The explicit closed-form solution for a linear differential equation put in the standard form is $x(t)= \frac{1}{\mu(t)} \times \left( \int \mu(t) \, Q(t) dt+ C \right)$, where $\mu (t)$ is given by $\mu(t) =  \exp\left(\int P(t)dt\right)$. 

The solution, then, for $x_t(t)$ at time $t$ in our IVP is given by the following expression:
\begin{align}\label{eq:sol_2cohort}
x_1(t)= \frac{1}{\mu(t)} \times \left\{ X_0\int \mu(t) \, g_0(t)\,  e^{-\delta t} dt+ C \right\},
\end{align}
where $C$ is an integration constant, and $\mu(t)$ is
\begin{align}
  \mu(t) &= \exp \left[ -\int (g_1- \delta) dt \right], \notag \\
  &= \exp \Bigl[ -(g_1-\delta)t \Bigr].
  \end{align}

Substituting the expression $g_0(t)$ (using Eq. \eqref{eq:g_c_explicit}), and the expression for $\mu(t)$ above into Eq. \eqref{eq:sol_2cohort} gives us
\begin{align} 
  x_1(t) &= \, {e^{ (g_1-\delta)t}} \times \Biggl\{ X_0\int \left(G_{T_{1}} - b_1\Delta^2_{0,1}\, e^{-2vt} \right)\,  e^{ -g_1 t}\,  dt + C\Biggr\} , \\
  &= \, X_0 \times \Biggl\{ \left(\frac{ b_1\Delta^2_{0,1}}{g_1 + 2v} \right)\,e^{-(\delta + 2v)t} \,-\, \left(\frac{ G_{T_{1}}}{g_1} \right)\, e^{-\delta t}  \Biggr\} \,+\, C  e^{(g_1-\delta )t} . \label{eq:sol_2cohort_a}
\end{align}
To find the integration constant $C$, we need to use the initial condition $x_1(t=0) =0$ which gives
\begin{align}
C = \, X_0 \left(\frac{ G_{T_1} }{g_1}    \,-\,   \frac{ b_1\Delta^2_{0,1}}{g_1 + 2v} \right).
\end{align}
then substituting the expression for $C$ back into Eq. \eqref{eq:sol_2cohort_a} gives the explicit solution for the population at time $t$,
\begin{align}
x_1(t) \,=\,  X_0 \, e^{(g_1-\delta )t} \cdot \left\{ \frac{ G_{T_1} }{g_1} \, \Bigl( 1 -  e^{-g_1 t} \Bigr)
\,\,-\,\, \frac{ b_1\Delta^2_{0,1}}{g_1 + 2v}  \Bigl(1 - e^{-(g_1  + 2v)t} \Bigr) \right\}.  
\end{align}

If we recall that the growth rate of individuals born in the current cohort/temperature regime, $g_1$, reflects the cohorts acclimation to the current temperature, then $g_1 = G_{T_1}$, which slightly simplifies the above expression. The solution for the total density of a population subject to a single temperature change is given by the sum of the densities of the two subpopulations or cohorts,
\begin{align} 
  x_0(t) \,&=\,  X_0 \, e^{-\delta t  } \\
  x_1(t) \,&=\,  X_0 \, e^{(g_1-\delta )t} \cdot \left\{ \Bigl( 1 -  e^{-g_1 t} \Bigr)
             \,\,-\,\, \frac{ b_1\Delta^2_{0,1}}{g_1 + 2v}  \Bigl(1 - e^{-(g_1  + 2v)t} \Bigr) \right\}, 
\end{align}
such that $x_{\mathrm{total}}(t)= x_0(t)+x_1(t)$. 

\subsection{Time-structured model under $N$ temperature shifts}
We can now extend the approach we used to solve the single temperature/two cohort model to solve the differential equations associated with the general case involving a population subject to $N$ distinct temperature regimes in a time-series. At time $t=0$ at the start of the current (or $N^\text{th}$) time period/temperature state, the initial densities of all previous cohorts, from zero to $N-1$, will be given by $X_{0,N}, X_{1,N},\ldots, X_{N-1, N}$, where $X_{c,N}$ is the initial density of the $c^\text{th}$ cohort during the $N^\text{th}$ temperature state or time period. (As per the assumptions of the TSM framework, the initial density of the current or  $N^\text{th}$ cohort, $X_{N,N}$ at $t=0$ should be zero.) Using the ODE system in Eq. \eqref{eq:TSM_ODE}, we can now define an Initial Value Problem (IVP) for this general TSM case: 
\begin{align} \label{eq:IVP_2temp}
  \frac{ d x_0}{dt} \quad\, &=\, - \delta x_0(t),   \notag \\
      &\,\, \vdots   \notag \\
  \frac{ d x_{N-1}}{dt} \, &=\, - \delta x_{N-1}(t),   \notag \\
  \frac{ d x_N}{dt} \quad &=\,  \sum_{c=0}^{N-1} g_c(t;T_N) \cdot x_c(t) \,\, + \,\,  \Bigl( g_N(T_N) - \delta_2 \Bigr) \cdot x_N(t), \\
  &\text{initial conditions:}\, x_c(0) =X_{c, N} \,\,\,(\forall c<N), \,\, \text{and} \,\, x_N(0)=0. \notag
\end{align}
 
Given that the explicit solution for all previous cohorts $x_c(t)$ (where $c<N$) is $x_c(t) = X_{_{c,N}} \, e^{-\delta t}$, we can rewrite the IVP in Eq \eqref{eq:IVP_2temp} above in the standard form as 
\begin{gather} \label{eq:IVP_general}
  \frac{ d x_N}{dt} \,-\,   \Big( g_N(T_N) - \delta \Big) \cdot x_N(t) \,=\,  \sum_{c=0}^{N-1} \Big(X_c \, e^{-\delta t}\cdot g_c(t; T_N)\Big),  \\
  \text{initial condition:} \, \quad x_N(0) = 0, 
\end{gather}
for which the explicit closed-form solution for $x_N(t)$ at time $t$ will be
\begin{align} \label{eq:integral_expression_N_cohorts}
x_N(t) &= \frac{1}{\mu(t)} \times \left\{ \int \mu(t) \,  \sum_{c=0}^{N-1} \Bigl(X_{c,N} \, g_c(t)\,  e^{-\delta t}\Bigr) dt+ C \right\},  \\
&= \frac{1}{\mu(t)} \times \left\{ \sum_{c=0}^{N-1} \,  X_{c,N} \, \int \Bigl(\mu(t) \,  g_c(t)\,  e^{-\delta t}\Bigr) dt+ C \right\}.
\end{align}
Here $C$ is an integration constant, and $\mu(t)$ is once again an integrating factor given by $\mu(t) = \exp \left\{ -(g_N-\delta)t \right\}$.

Substituting in the expression $g_c(t)= G_N - b_N\Delta_{c, N}^2\, e^{-2vt}$ (Eq.\eqref{eq:g_c_explicit}), where $b_N$ is a parameter constant during the current $N^{th}$ temperature state, and the parameter constant $\Delta_{c,N}$ is the difference between the current ambient temperature, $T_N$ and the \textit{initial} acclimated temperature of the $c^{th}$ cohort, $T_{A_c}^{\,\circ}$, during the current $N^{th}$ temperature state, gives us 
\begin{align} \label{eq:x_N_sol1}
  x_N(t) &= \, {e^{ (g_N-\delta)t}} \times \Biggl\{ \sum_{c=0}^{N-1} \, X_{c,N}\int \left(G_{T_N} - b\Delta^2_{c,N}\, e^{-2vt} \right)\,  e^{ -g_N t}\,  dt + C\Biggr\} , \notag\\
  &= \, \sum_{c=0}^{N-1} \, X_{c,N} \times  \Biggl\{ \left(\frac{ b\Delta^2_{c,N}}{g_N + 2v} \right)\,e^{-(\delta + 2v)t} \,-\, \left(\frac{ G_{T_N}}{g_N} \right)\, e^{-\delta t}  \Biggr\} \,+\, C  e^{(g_N-\delta )t} .
\end{align}
To solve for the integration constant $C$ we need to use the initial condition $x_N(t=0) =0$
\begin{align}
C = \, \sum_{c=0}^{N-1} X_{c,N} \left(\frac{ G_{T_N} }{g_N}    \,-\,   \frac{ b\Delta^2_{c,N}}{g_N + 2v} \right).
\end{align}

Substituting the expression for $C$ back into Eq. \eqref{eq:x_N_sol1}, and recalling that $g_N = G_{T_N}$, gives the explicit solution for each of the cohorts in the population at time $t$ as
\begin{align} \label{eq:x_N_sol2}
  x_c(t) \,\,&=\,  X_{c,N}\,\, e^{-\delta t} \quad\quad(\text{for all} \,\, 0\leq c < N) \\
x_N(t) \,&=\, \, e^{(G_{T_N}-\delta )t} \times \sum_{c=0}^{N-1}  \left\{  \,X_{c,N} \, \Bigl( 1 -  e^{-G_{T_N} t} \Bigr) \right. \notag\\
&\quad\quad\quad\quad\quad\quad\quad \left. \,\,-\,\, X_{c,N} \, \frac{ b_N\Delta^2_{_{c,N}}}{G_{T_N} + 2v}  \Bigl(1 - e^{-(G_{T_N}  + 2v)t} \Bigr) \right\}.  
\end{align}

\subsubsection{Parameter values for solving TSMs}
In the closed form solution to the TSM differential equation system in Eq. \eqref{eq:x_N_sol2}, there are numerous parameter values that are determined by past time periods and temperature states; specifically, the initial densities of previous cohorts, $X_{c,N}$, and the initial acclimation temperature, $T^{\,\circ}_{A_c}$, at the start of the current period when $t=0$. Since these initial parameters are determined by previous temperature cycles, we can find the expressions for each of them by using an iterative approach.

\begin{tabular}{ll}  
  \toprule
  \multicolumn{2}{c}{\textbf{Table 1: Summary of variables and parameters}} \\
  \midrule
  Parameters         & Description \\
  \midrule
  $N$               & Current temperature state, as well as total number of\\
                    & states in time series ($1\leq N$)                           \\
  $c$               & Cohort value, where $c\in\{0,1\ldots,N\}$  \\
  $T_N$ (or $T_k$)  & Temperature of the current $N^{th}$ (or $k^{th}$)  state in time series  \\
  $T_{A_c}^{\circ}$ & acclimated temperature of the $c^{th}$ cohort at beginning of \\
                    & current or given temperature state \\
  $b_k$             & Parameter constant specific to temperature $T_k$ \\
                    & for determining curvature of parabola for $g_c$     \\
  $G_{T_N}$         & Optimum growth rate at the current temperature \\
  $v$               & Speed of temperature acclimation \\
  $X_{c,N}$         & Initial density value of $c^{th}$ cohort for current $N^{th}$ state \\ 
  $\tau_k$          & Time spent in the previous $k^{th}$ temperature state,   \\
                    & where $k\in\{1,\ldots, N-1\}$           \\
  \midrule
  Variables         & Description \\
  \midrule
  $x_c (t)$       & Density of the $c^{th}$ cohort at time $t$\\
  $t$         & Time in the current ($N^{th}$) temperature state   \\
  \midrule
  Parameter functions    & Description \\
  \midrule
  $\Delta_{c,k}$         & Difference between current temperature $T_k$ and initial \\
                         & acclimated temperature, $T_{A_c}^{\circ}$: $\Delta_{c,k}= T_k - T_{A_c}^{\circ}$ \\
  $g_c$                  & Contributions from the $c^{th}$ cohort to the growth \\
                         & rate of current ($N^{th}$) cohort  \\
  \bottomrule
\end{tabular}

\paragraph{Initial density parameters} The values of all the initial density parameters, $X_{c,N}$, appearing above in solution Eq. \eqref{eq:x_N_sol2} are themselves determined by the history of population growth prior to the current period -- that is, before time $t=0$. If we know the history of temperature changes that the population experienced previously, and the time duration of each previous temperature state, $\tau$, then we can directly compute the initial parameter values for the current period, $X_{c,N}$, by iteratively solving Eq. \eqref{eq:x_N_sol2}, starting with $N=1$ and proceeding until we reach the value of $N-1$. 

When using the above approach, we first note $X_{0,1}$ is the constant representing the initial population size at the start of the time series, that is, before the first temperature state $T_1$ occurs. For each of the other parameters, the value of $X_{c,N}$ is simple to determine when one realizes that each $c$ cohort was produced in the $c^\text{th}$ temperature/time period as per Eq. \eqref{eq:x_N_sol2} above, and that each prior cohort subsequently decays after the temperature/time period it was created in passes, as is given by the expression $\exp\left\{-\delta \sum_{i=c+1}^{N-1} \tau_i\right\}$ (where $\tau_i$ is the time duration of the $i^{th}$ temperature state). For the initial parameter value of the zeroth cohort, $X_{0,N}$, this gives us
\begin{align}
  X_{0,N} = X_{0,1} \, e^{-\delta \sum_{1}^{N-1} \tau_i}
\end{align}
For all other previous cohorts $c$, where $1 \leq c < N$, the initial parameter value $X_{c,N}$ (which is solvable by iterative substitution) becomes 

  \begin{align}
    X_{c,N} \,&=\,  e^{\left(G_{T_N}\tau_c -\delta \sum_{c}^{N-1} \tau_i  \right)} \times 
    \sum_{k=0}^{c-1}  \left\{  \,X_{k,c} \, \Bigl( 1 -  e^{-G_{T_c} \tau_c} \Bigr) \,\,-\,\, X_{k,c} \, \frac{ b\Delta^2_{k,c}}{G_{T_c} + 2v}  \Bigl(1 - e^{-(G_{T_c}  + 2v)\tau_c} \Bigr) \right\}.   
\end{align}

\paragraph{Initial acclimated temperature parameters} Recall from Eq. \eqref{eq:T_A} that $\Delta_{c,N}$ is equal to $T_N - T^{\,\circ}_{A_c}$, that is, the difference between whatever the current ambient temperature is, $T_N$, and the temperature that the $c^\text{th}$ cohort was initially acclimated to at the beginning of the current time period (i.e., at time $t=0$). This means that at time $t=0$ we will need initial parameter values $T^{\,\circ}_{A_c}$ for each  of the previous $c$ cohorts at the start of the current time period, $T_{A_c}(0)= T_{A_c}^{\,\circ}$. The expressions for these initial parameter values can be found, like for the initial densities, by iteratively solving the expression for $T_{A_c}$ given in Eq. \eqref{eq:T_A} for each of the previous temperature states, starting at $T_1$, and substituting the solution at the end of the temperature period into the equation for the subsequent temperature state (see Appendix A for details). The parameter values for $T^{\,\circ}_{A_c}$, then, are as follows:
\begin{align} \label{eq:T_Acclim_0_first}
  T_{A_{N-1}}^{\,\circ} &=  T_{N-1}  &\text{for $c=N-1$}, \\
  T_{A_{N-2}}^{\,\circ} &=  T_{N-1} - (T_{N-1} -T_{N-2})e^{-v\tau_{N-1}} &\text{for $c=N-2$},
\end{align}
And for all $c < N-2$,
\begin{align} \label{eq:T_Acclim_0_second}
  T_{A_c}^{\,\circ}  = T_c\times \exp\left[-v\sum^{N-1}_{i=c+1} \tau_i \right] \,\,+\,\, \sum^{N-2}_{i=c+1} T_i \, \Bigl(1- e^{-v\tau_i} \Bigr)\times \exp{ \left[-v\sum^{N-1}_{j=c+2} \tau_j\right] } \quad\quad \notag \\
    +T{_{N-1}}\Bigl(1- e^{-v\tau_{_{N-1}}} \Bigr). \quad\quad\quad\quad
\end{align}

In Box 1 below, we present and summarize the explicit solution to differential equations describing exponential population growth in time-structured models. If we wish to simplify and have just a single explicit expression for the total population density, we can sum up the cohort densities in Box 1 to get a single expression for $x_{\text{total}}(t)$.

\begin{tcolorbox}[breakable, enhanced]
  \centerline{\textbf{Box 1}} 
  \centerline{\textbf{A general time-structured model for population growth}} 
    In a time series where the environment cycles through $N$ discrete environmental/temperature phases ($N\in\mathbb{Z}^+$), the total density $x_{\text{total}}(t)$ of a population made of $N+1$ cohorts at a given time $t$ during the $N^\text{th}$ environmental/temperature regime can be written as 
    \begin{align}
      x_{\text{total}}(t) = x_0(t) + x_1(t) + \cdots + x_c(t) + \cdots + x_N(t), 
    \end{align}
    where the densities of all the cohorts $x_0(t), \ldots , x_c(t), \ldots , x_N(t)$ at time $t$ are given by
    \begin{align}
      x_0(t)  \, &= \, X_{_{0,N}} \,\, e^{ -\delta t } \,=\, X_{0,1}\,\, e^{-\delta \left(t+\sum_1^{N-1} \tau_i \right) },   \\
      x_c(t) \, &=\,  X_{c,N}\,\, e^{-\delta t} \quad\quad (\text{for all} \,\, 0 < c < N), \\
      x_N(t) \, &=\, \, e^{(G_{T_N}-\delta )t} \times \sum_{c=0}^{N-1}  \left\{  \,X_{c,N} \, \Bigl( 1 -  e^{-G_{T_N} t} \Bigr) \right. \quad \notag\\
      \quad&\quad\quad\quad\quad\quad\quad\quad \left. \,\,-\,\, X_{c,N} \, \frac{ b_N\Delta^2_{_{c,N}}}{G_{T_N} + 2v}  \Bigl(1 - e^{-(G_{T_N}  + 2v)t} \Bigr) \right\}. 
    \end{align}
    The initial value $X_{0,N}$ is equal to $X_{0,1}\,\, e^{-\delta \sum_{1}^{N-1} \tau_i}$, where $X_{0,1}$ is the initial population density at the beginning of the time series (i.e., at the start of the first environmental state). The values for all other initial cohort densities, $X_{c,N}$, at the start of the current period are given by 
    \begin{align}
      X_{c,N} \,&=\,  e^{\left(G_{T_N}\tau_c -\delta \sum_{c}^{N-1} \tau_i  \right)} 
       \times  \sum_{k=0}^{c-1}  \left\{  \,X_{k,c} \, \Bigl( 1 -  e^{-G_{T_c} \tau_c} \Bigr) \right.  \notag \\ 
       &\quad\quad\quad\quad\quad\quad\quad\quad\quad\quad\quad\quad \left. \,\,-\,\, X_{k,c} \, \frac{ b_c\Delta^2_{k,c}}{G_{T_c} + 2v}  \Bigl(1 - e^{-(G_{T_c}  + 2v)\tau_c} \Bigr) \right\}, 
    \end{align}
    for each $c\in\{1, \ldots , (N-1)\}$.

    In the expression for initial densities, $X_{k,c}$, the parameter constant $\Delta_{k,c}$ now represents the difference between the initial acclimated temperature of a cohort $k$ that preceded one of the previous $c$ temperature states, $\Delta_{k,c}= T_c -T_{A_k}^{\circ}$. It is the $\Delta$ constant associated with past time periods or temperature states.

\end{tcolorbox}

\section{TSMs under limiting cases}
The general time-structured model derived above can, under specific system assumptions or requirements, give rise to different forms of time-structured models. Here we briefly describe two limiting cases.
\subsection{Homogeneous populations}
An important limiting case for TSMs corresponds to a scenario whereby all progeny produced under a given temperature regime perfectly inherit the acclimation temperature of their parents at the moment of birth. Furthermore, the progeny exhibit phenotypic plasticity so that they change their temperature acclimation, and consequently their reproductive output, at the same rate as their parents. These are the the demographic assumptions that are implicit in the model used by Kremer et al. \citep{kremer.fey.ea_18} to experimentally study the basis of phenotypic plasticity under fluctuating temperature.

Under these assumptions, all individuals regardless of their cohort generation, are `synchronized' in both their temperature acclimation and reproductive output rates. Since all cohorts are now identical, there is effectively only a single cohort composed of indistinguishable individuals. In the TSM framework, this means that the reproduction of individuals in the current ($N^\text{th}$) period, $g_N$, will be identical to the reproductive output of all other previous $c$ generational cohorts, $g_c$: $g_N = g_c, (\forall c\leq N)$. The expression in Eq. \eqref{eq:integral_expression_N_cohorts} representing the solution to the TSM differential equations now changes and becomes the integral expression shown below. 

Under these assumptions the size of the $N^\text{th}$ cohort according to Eq. \eqref{eq:integral_expression_N_cohorts} is 
\begin{align}
  x_N(t) &= \, \frac{1}{\mu(t)} \times \left\{ \sum_{c=0}^{N-1} \,  X_{c,N} \, \int \Bigl(\mu(t) \,  g_N(t)\,  e^{-\delta t}\Bigr) dt+ C \right\}, \notag \\
         &= \, {e^{ (g_N-\delta)t}} \times \Biggl\{ \sum_{c=0}^{N-1} \, X_{c,N}\int \left(G_{T_N} - b\Delta^2_{c,N}\, e^{-2vt} \right)\,  e^{ -g_N t}\,  dt + C\Biggr\} .
  \end{align}
  (Recall from earlier that $\mu(t) =\exp[ -(g_N -\delta)t]$.) 
  
  Taking the solution to the above integral expression and applying it iteratively to all previous $N-1$ cohorts starting with the first cohort will allow us to calculate the density of each cohort (see Appendix B). When all cohorts are summed together at time $t$, we get $x_{\mathrm{total}}= x_0 + x_1 + \cdots + x_N$, the total density for what is now a non time-structured model (nTSM), whose closed form solution is presented in Box 2.

\begin{tcolorbox}[breakable, enhanced]
    \centerline{\textbf{Box 2}} 
    \centerline{\textbf{Non time-structured model for homogeneous populations}}
    If a homogeneous population (where all individuals are indistinguishable and acclimate in unison) has been subjected to $N$ distinct temperature or environmental states over time, then the total density of the population $x_{\mathrm{\,total}}(t) $ at any given time $t$ during the current $N^{th}$ temperature state will be
    \begin{align} \label{eq:homogeneous_pop}
      x_{\mathrm{\,total}}(t) &\,=\,  X_0 \,\, e^{(G_{T_N} -\delta )t} \times \exp\left\{ \, \frac{b_c\Delta_N^2}{2v} \left( e^{-2vt}  -  1 \right) \right. \notag \\ 
      & \left. \quad\quad\quad\quad\quad + \sum_{c=1}^{N-1} \left[ (G_{T_c} -\delta )\tau_c +  \frac{b_c\Delta_c^2}{2v} \left( e^{-2v\tau_c}  -  1 \right) \right] \right\},
    \end{align}
    where $X_0$ is the initial population density at the beginning of the time series (i.e., the start of the $1^\text{st}$ time period), and $\Delta_c = T_c - T^{\circ}_{A_{\bullet}}$, where $T^{\circ}_{A_{\bullet}}$ is initial acclimation temperature of the total population at the beginning of the $c^{th}$ temperature state. 
  \end{tcolorbox}

  Alternatively, for this special limiting case involving a homogenous population, the expression for the total density can also be found in more straightforward manner by directly solving the separable linear differential equation describing population growth (see Appendix B).

  \subsection{Populations exhibiting no phenotypic plasticity}
  Another limiting case involves the scenario whereby a population has distinct cohorts, but where all the individuals are unable to track or progressively adapt to changes in environmental conditions. The lack of adaptability or phenotypic plasticity among individuals in all cohorts is simple to implement by setting the rate or speed of acclimation to zero, $v=0$. The solution for such a scenario is given in Box 3.

\begin{tcolorbox}[breakable, enhanced]
    \centerline{\textbf{Box 3}} 
    \centerline{\textbf{Time-structured model with no phenotypic plasticity}}
    For a  population structured by $N$ distinct temperature or environmental states and where individuals in each cohort have no phenotypic plasticity and thus are unable to acclimate to temperature (acclimation velocity $v=0$), then the total population  density at time $t$ during the current $N^\text{th}$ temperature state will be 
    \begin{align}
      x_{\text{total}}(t) = x_0(t) +  \cdots + x_c(t) + \cdots + x_N(t), 
    \end{align}
    where each of the cohort densities are
    \begin{align}
      x_0(t)  \,&= \, X_{_{0,N}} \,\, e^{ -\delta t } \,=\, X_{0,1}\,\, e^{-\delta \left(t+\sum_1^{N-1} \tau_i \right) },   \\
      &\,\vdots  \notag\\
      x_c(t) \,&=\,  X_{_{c,N}} \,\, e^{ -\delta t },  \\
      &\,\vdots  \notag \\
      x_N(t) \,&=\, \, e^{(G_{T_N}-\delta )t} \times \sum_{c=0}^{N-1}  \left\{  \,X_{c,N} \,\frac{ G_{T_N} - b_N\Delta_{c,N}^2 }{G_{T_N} } \Bigl( 1 -  e^{-G_{T_N} \tau_c} \Bigr)  \right\}.  
      \end{align}
    Each cohort's $\Delta_c$ value will be the same regardless of temperature/time period it was produced in (each individual's acclimation temperature never changes). Parameter $X_{0,1}$ is the initial population density at the beginning of the time series (i.e., at the start of the 1st environmental state). The values for all other initial cohort densities, $X_{c,N}$, at the start of the current period are given by 
    \begin{align}
      X_{c,N} \,&=\,  e^{\left(G_{T_N}\tau_c -\delta \sum_{c}^{N-1} \tau_i  \right)} \times  \sum_{k=0}^{c-1}  \left\{  \,X_{k,c} \,\frac{ G_{T_c} - b_c\Delta_{k,c}^2 }{G_{T_c} } \Bigl( 1 -  e^{-G_{T_c} \tau_c} \Bigr)  \right\}.  
    \end{align}
  \end{tcolorbox}

\section{Case studies of various TSM and nTSM growth models}
In this section, we illustrate the various outcomes that are possible when using the different scenarios and growth models derived above. These scenarios offer a proof-of-concept for how the different assumptions underlying the various growth models can produce drastically different outcomes. For all scenarios, the population was subjected to the same series of twenty temperature fluctuations ($N=20$) modeled using a square wave function, with the temperature oscillating between $14^{\circ} \textrm{C}$ and $30^{\circ} \textrm{C}$, and where the duration of all temperature states is equal ($\tau = 1$ day). For convenience, we used the generation rates that were provided by Kremer et al.\citep{kremer.fey.ea_18} in Table B1 of Appendix B (see Appendix C of this paper for all parameters used in time series). 

\begin{figure}
  \centering
  \includegraphics[width=0.75\textheight]{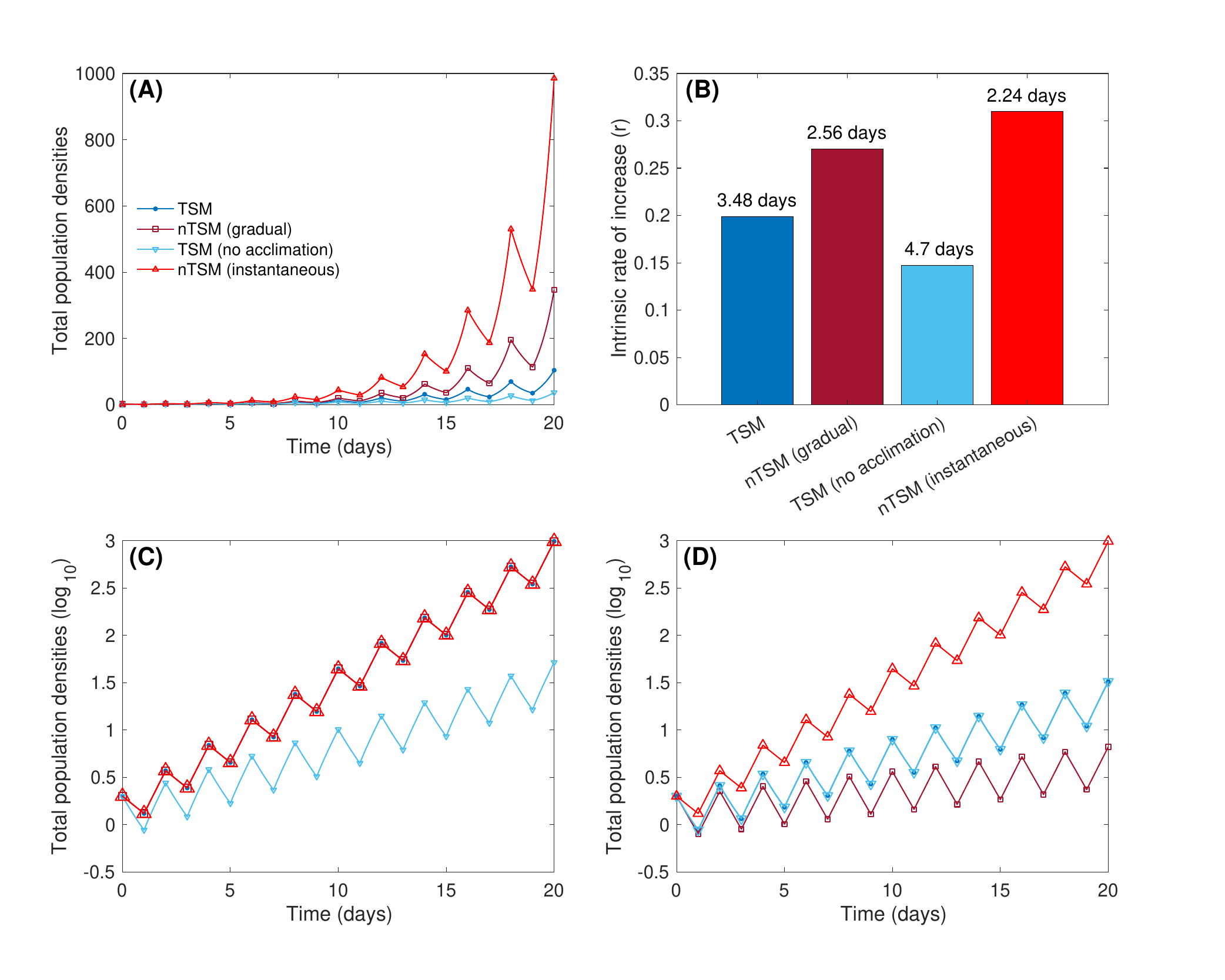}
  \caption{\textbf{Large variation in population densities observable between TSMs and nTSMs}. (A) Time series of densities for a basic TSM (dark blue filled circles) and an nTSM with gradual acclimation (homogeneous population model; burgundy squares). Both are plotted alongside, for comparison, a time series for a TSM with no acclimation ability (light blue upside down triangles) and an nTSM with instantaneous acclimation (red triangles). (B) Bar plots showing average intrinsic growth rates ($r$) across full time series. Values for doubling time (in days) are provided above each bar. (C) TSM and nTSM (gradual) time series ($\log_{10}$) when $v\to\infty$. Both time series converge with the nTSM (instantaneous acclimation). (C) TSM and nTSM (gradual acclimation) time series ($\log_{10}$) when $v\to0$. The basic TSM (dark blue filled circles) now converges to the time series represented by the TSM (no acclimation) growth model (light blue upside down triangles).}
  \label{Fig:fig2}
\end{figure}

We plotted the time series for a basic TSM and an nTSM (homogeneous growth model), both of which assumed an acclimation rate of $v=0.35$ (filled circles and squares, respectively; Fig 2A). We also plotted for comparison two reference time series representing two limiting cases: a TSM growth model for a population with no plasticity ($v=0$) and an nTSM (homogeneous model) for a population with instantaneous acclimation ($v\to\infty$; Fig 2A). Thus we have examples of two TSM growth models (a basic TSM represented by Eq. \eqref{eq:x_N_sol2}, and a TSM with no acclimation), as well as two nTSMs (a basic homogeneous model, Eq. \eqref{eq:homogeneous_pop}, and a homogeneous growth model with instantaneous acclimation).

For the given set of parameters (see Appendix C), all four exponential growth models diverge significantly in outcomes. Unsurprisingly, the nTSM (homogeneous model) with rapid acclimation is able to achieve maximum growth by instantaneously adapting to each new temperature regime (red curve, Fig 2A). On the other hand, the TSM with no plasticity suffers the lowest growth rate possible because it simulates a population composed of multiple cohorts at maximum distance from their growth optimum (blue curve, Fig 2A). The basic TSM and nTSM growth models lie in between these outcomes, with the simple nTSM (homogeneous model similar to the one studied by Kremer et al. \citep{kremer.fey.ea_18}) having a density three times that of the basic TSM that exhibits the effects of past temperature fluctuations. These differences can be summarized by comparing the intrinsic growth rates as well as the doubling time between our nTSM and TSM models (Fig 2B). Here, despite the limited length of the time series, both the intrinsic growth rates and the doubling times varied by as much as a factor of 2 across the scenarios. 

If we take our basic TSM and nTSM and allow $v\to\infty$, then both of their time series will converge with the time series for the instantaneous acclimation model, as would be expected (Fig 2C). If, however, we allow $v\to 0$ for both models, then the TSM time series will instead converge towards the TSM without acclimation (again, as would be expected), while the time series for the homogeneous nTSM population model in this case study drops below that of all other models as it lacks the variation in temperature acclimation provided by the presence of multiple cohorts, and as a result, the population is less buffered from the effects of maladapted growth (Fig 2D).  

\section{Discussion}
We have shown that nTSM approaches used to quantify the dynamics of populations exposed to temporal environmental variation yield potentially large errors because they induce a kind of memory loss via their tacit assumption that all individuals respond identically to environmental change despite their differential exposure to environmental conditions. This assumption may hold under extremely limited conditions, such as in very short experiments or for organisms with extremely limited lifespans, but in most situations one must turn to TSMs in order to account for the emergence of cryptic population structure under environmental fluctuations, and its resulting historical effects on population dynamics. Indeed, alternatives to TSMs such as statistical frameworks based on autoregressive and Bayesian hierarchical models built explicitly to quantify the kinds of historical effects that arise under fluctuating environments \citep{ogle_ea_15} will suffer from the same forms of ecological memory loss as nTSMs when applied to seemingly homogeneous populations where individuals cannot be tracked over time (e.g., phytoplankton or bacteria) because these methods do not account for the cryptic population structure that emerges when organisms are exposed to variable environmental conditions.

TSMs are a flexible tool that can be instantiated to simulate growth under many different scenarios. For example, we initially assumed that the death rate $\delta_c$ was the same for each $c$ cohort, yet solutions for population densities could be found for cases when each cohort's death rate changes as a function of temperature. Many other model variations and assumptions are also possible, some of which lend themselves to being solved explicitly depending on the nature of the integrals. 

Although explicit solutions for the TSMs presented in this paper can be obtained directly using closed form expressions, the number of inputs (i.e., initial values) required to solve the ODE system (and thus the number of cohorts that need to be tracked) will grow as the total number of environmental states increases. This, however, should not present any serious computational problems, particularly for studies that range over normal ecological or evolutionary time spans. If one wishes to compute how population (and cohort) density changes over a set number of environmental states or fluctuations, then pre-allocating matrices for storing densities based on the maximum number of cohorts should allow for efficient computation of any and all cohort densities at any point in time.

The general closed form solution presented here represents a formal extension of simple exponential growth from classical population biology that can be used to study the impacts of environmental variation on population growth and persistence, as well as the links between evolutionary and phenotypic plasticity \citep{kremer.fey.ea_18,fey_ea_21}. This is underscored by the ease with which the TSM framework is able to \textit{directly} and immediately recover results observed in recent studies without the need for extensive simulation work. In fact, the robustness of the TSM framework ensures that, often, only minor modifications will be needed for accommodating the requirements of current experimental models and studies.

For instance, in  our solutions, we used a specific function to relate how a population's reproductive rate $g$ will be affected by differences between the ambient and acclimated temperatures $g_c(t) = G_{T_N} - b_N\,\left(T_N - T_{{A}_c} (t)\right)^2$ (Eq. \eqref{eq:g_c}). We used a smooth parabolic function, $\left(T_N - T_{{A}_c } (t)\right)^2$, because it was the simplest functional relationship possible, requiring only minimal assumptions while at the same time being able to capture the phenomenological nature of change. However, the choice of this or other simple functional relationships that have been assumed in current models or experiments is unlikely to have a dramatic impact on the ODE solutions; one need only replace the expression for squared temperature difference with a general undefined function $f\left(T_{{A}_c } (t)\right)$ when solving the ODE system.

As an example, take the Kremer et al. \citep{kremer.fey.ea_18} study which used a \textit{non} time-structured (homogeneous) population growth model to provide the baseline expectations when experimentally studying the effects of gradual temperature acclimation in microorganisms. The phenomenological growth equation that they used to capture plasticity under varying temperatures is similar to our function in Eq. \eqref{eq:g_c}, $g_c(t) = G_{T_N} - b_N\,f(T_{{A}_c}(t))$. However, while in our equation the function $f$ is defined using a parabolic curve $f\left(T_{{A}_c} (t)\right) = \left(T_N - T_{{A}_c}(t) \right)^2$, Kremer et al. \citep{kremer.fey.ea_18} defined this function as the absolute value of the difference between current and acclimated temperature: $f\left(T_{{A}_c}(t) \right) = \left|T_N - T_{{A}_c} (t)\right|$.

If we wished to use our general TSM equations to recover the results that Kremer et al. \citep{kremer.fey.ea_18} obtained through numerical simulations of homogeneous population growth, then in addition to using the homogeneous population equation derived above in Box 2, we would need to replace the parabolic equation used for the growth rate with Kremer et al.'s function to obtain $g_c(t) = G_{T_N} - b_N \cdot f\left(T_{{A}_c} (t)\right) = G_{T_N} - b_N\left| \Delta_{c,N} \,\right|\, e^{-vt}$, where $\Delta_c$ is the parameter constant $\Delta_c = T_N- T^{\,\circ}_{A_c}$.

The closed form solutions for TSMs represent solutions for cases where baseline growth in non-varying environments is likely to be exponential in nature, as one might expect in experimental or chemostat systems or in large natural systems where population densities can be considered sufficiently far away from equilibrium. Dealing with near equilibrium cases or situations where nonlinearities make solutions to ODEs difficult to solve may require more simulation-based approaches. In situations where closed form solutions may not be attainable, one can simply ``solve'' the relevant TSM numerically using existing numerical schemes (e.g., Runge-Kutta method). Overall, TSMs represent an important new tool for studying the effects of environmental variation on population growth, size, and persistence in an era of global change.

\section*{Acknowledgements}
We acknowledge support from the National Science Foundation (OCE-2048894 and OCE-2308605).

\section*{Authors' contributions}
PP developed the conceptual approach, derived the quantitative methods, plotted the figures, and wrote the manuscript. TG contributed to/helped edit the manuscript, and helped to edit the figure plots.

\bibliographystyle{ecology_letters}
\bibliography{pillai_gouhier_TSM_2023.bib}

\newpage{}

\section*{Appendix A: }
Recall from Eq. \eqref{eq:T_A} that $\Delta_{c,N}$ is a constant equal to $T_N - T^{\,\circ}_{A_c}$, which is, the difference between whatever the current ambient temperature is, $T_N$, and the temperature that the $c^{th}$ cohort was initially acclimated to at the beginning of the current time period, $T^{\,\circ}_{A_c}$ (i.e., at time $t=0$). This means that  we will need to know the initial parameter values $T^{\,\circ}_{A_c}$ for each  of the previous $c$ cohorts at at time $t=0$, that is, at the start of the current time period, $T_{A_c}(0)= T_{A_c}^{\,\circ}$. The expressions for these initial parameter values can be found, like for the initial densities, by iteratively solving the expression for $T_{A_c}$ given in Eq. \eqref{eq:T_A} for each of the previous temperature states, starting at $T_1$, and substituting the solution at the end of the temperature period into the equation for the subsequent temperature state. 

Consider, for example, the acclimation temperature of the zeroth cohort. Let us assume that before the time series has begun the initial population was originally acclimated to temperature of $T_0$. Then, at the end of the first time period of duration $\tau_1$, defined by the temperature state $T_1$,  we have 
\begin{align} \label{eq:A1}
   \text{End of first period:} \quad T_{A_0} (\tau_1) = T_1 - \left(T_1-T_0\right)\,e^{-v\tau_1}. \quad\quad
\end{align}
The acclimation temperature of the zeroth cohort at the end of the first time period serves as the initial acclimation temperature at the start of the second period (temperature $T_2$). This means the acclimated temperature at the start of the second temperature state/period  ($T_2$) is $T_{A_0}^{\,\circ}=T_{A_0} (\tau_1)$. 

At the end of the second period the acclimated temperature of the zeroth cohort will then be as follows,
\begin{align} 
  \text{End of second period:} & \notag \\
  T_{A_0} (\tau_1+\tau_2) &= T_1 - \left( T_2 - T_{A_0}^{\,\circ} \right)\,e^{-v\tau_2}, \notag \\
  &= T_2 - \Bigl( T_2 - \left[  T_1 - \left(T_1-T_0\right)\,e^{-v\tau_1} \right] \Bigr)\,e^{-v\tau_2} \notag \\
  &= T_2 - T_2\,e^{-v\tau_2} + T_1 \,e^{-v\tau_1}  -  T_1 \,e^{-v(\tau_1+\tau_2)}  + T_0 \,e^{-v(\tau_1+\tau_2)} \notag \\
  &=  T_0 \,e^{-v(\tau_1+\tau_2)} +T_1 \,e^{-v(\tau_2)}(1-e^{-v\tau_1})  + T_2 (1-e^{-v\tau_2}) .
\end{align}
(Note above the substitution of the expression for the first period into the expression for the second.) Similarly, at the end of the third temperature interval (or beginning of the fourth) we have 
\begin{align} 
  \text{End of third period:}& \notag \\
   T_{A_0} (\tau_1+\tau_2+\tau_3) &=  T_0 \,e^{-v(\tau_1+\tau_2+\tau_3)} +T_1 \,e^{-v(\tau_2+\tau_3)}(1-e^{-v\tau_1})  \notag \\
   & \quad\quad\quad\quad + T_2\,e^{-v\tau_3}(1-e^{-v\tau_2}) + T_3 (1 - e^{-v\tau_3} ).
\end{align}

Using sequential substitution, we can establish the general expression for the zeroth cohort's acclimated temperature at the beginning of the current (or $N^{th}$) temperature state -- or alternatively, at the end of the $N-1$ temperature state, $T_{A_0}^{\,\circ} = T_{A_0} (t=0)$:

acclimated temperature at start of $N^{th}$ period,
\begin{align} \label{eq:Append_T_acclim_zeroth}
   T_{A_0} (t=0) &=  T_0 \,e^{-v \sum_1^{N-1} \tau_i} + \sum_{i=1}^{N-2} T_i (1-e^{-v\tau_i})\,e^{-v \sum_{2}^{N-1}\tau_j}  + T_{N-1}\,(1 - e^{-v\tau_{N-1}} ).
\end{align}

We can generalize the above iterative derivation for for all other cohorts $c$, such that $c < N$. For $c=N-1$ and $c=N-2$ the expressions are straight forward from the initial expressions determined above:
\begin{align} \label{eq:Append_A}
  T_{A_{N-1}}^{\,\circ} &=  T_{N-1}  &\text{for $c=N-1$}, \\
  T_{A_{N-2}}^{\,\circ} &=  T_{N-1} - (T_{N-1} -T_{N-2})e^{-v\tau_{N-1}} &\text{for $c=N-2$}.
\end{align}
When $c < N-2$ the expression is a generalization of Eq. \eqref{eq:Append_T_acclim_zeroth},
\begin{align} 
  T_{A_c}^{\,\circ}  = T_c\times \exp\left\{-v\sum^{N-1}_{i=c+1} \tau_i \right\} \,\,+\,\, \sum^{N-2}_{i=c+1} \left[ T_i \, \Bigl(1- e^{-v\tau_i} \Bigr)\times \exp{ \left\{-v\sum^{N-1}_{j=i+1} \tau_j\right\} } \right] \quad\quad \notag \\
    +T{_{N-1}}\Bigl(1- e^{-v\tau_{_{N-1}}} \Bigr). \quad\quad\quad\quad
\end{align}

\section*{Appendix B: Homogeneous population case}
We demonstrate here how to derive the expression for the special case involving exponential growth in a homogeneous population (where all cohorts are identical) when the population is subject to a temporally varying environment (Eq. \eqref{eq:homogeneous_pop} in the main text). We first derive the expression by solving the differential equation associated with the TSM framework. Then as a comparison, we will derive the same expression more directly by solving the simple separable first order differential equation describing the same scenario.

\subsection*{Derivation using the TSM framework}
Recall the differential equation for TSM growth in Eq. \eqref{eq:IVP_general} in the main text:
\begin{gather}
  \frac{ d x_N}{dt} \,-\,   \Big( g_N(T_C) - \delta \Big) \cdot x_N(t) \,=\,  \sum_{c=0}^{N-1} \Big(X_c \, e^{-\delta t}\cdot g_c(t; T_C)\Big),  \\
  \text{initial condition:} \, \quad x_N(0) = 0, 
\end{gather}
for which the explicit closed-form solution for $x_N(t)$ at time $t$ will be in the form
\begin{align} 
x_N(t) &= \frac{1}{\mu(t)} \times \left\{ \int \mu(t) \,  \sum_{c=0}^{N-1} \Bigl(X_{c,N} \, g_c(t)\,  e^{-\delta t}\Bigr) dt+ C \right\},
\end{align}
where $C$ is an integration constant, and $\mu(t)$ is an integrating factor given by $\mu(t) =  \exp\left(\int (g_N-\delta)dt\right)$. Now since all cohorts are identical, we must first note that the growth rate of the current cohort, $g_N$, will be identical to the growth rates of all previous cohorts $g_c$ such that all cohorts will be indistinguishable, $g_N = g_c,\forall c$. Furthermore, since the current cohort is identical to all previous ones it will vary through time with the rest of the population $g_N(t) = G_{T_N} - b\Delta_{N}^2 e^{-2vt}$, where $\Delta_{N} = T_N - T^{\circ}_{A_\bullet}$, and $ T^{\circ}_{A_\bullet}$ is the initial acclimated temperature for the \textit{total} population at the start of the current temperature state. 

Since $g_N (t)$ now changes with time, the expression for the integration parameter $\mu(t)$ will be different from before. If we take $h(t)$ to be the antiderivative of $g_N$, such that $d h(t)/ dt = g_N$, then for the integration parameter we have
\begin{align}
  \mu(t) &= \exp \left[ -\int (g_N(t) - \delta) dt \right], \notag \\
  &= \exp \left[ -(h(t) - \delta t) \right].
\end{align}
This now gives
\begin{align} 
  x_N(t) &= e^{h(t)-\delta t} \times \left\{ \int   \,  \sum_{c=0}^{N-1} \Bigl(X_{c,N} \, g_N(t)\,  e^{-h(t)} \Bigr) dt+ C \right\},
  \end{align}
Once we recall that $\mathrm{d} h = g_N\,\mathrm{d}t$, then the integral above becomes straightforward to solve
  \begin{align} 
    x_N(t) &= e^{h(t)-\delta t} \times \left\{ \sum_{c=0}^{N-1} X_{c,N} \times \int  \Bigl( \,  e^{-h(t)} d h  \Bigr) + C \right\},\notag \\
    &=   C\,e^{h(t)-\delta t} - e^{-\delta t} \sum_{c=0}^{N-1} X_{c,N} .   
  \end{align}
Noting that $h=G_{T_N} + \frac{ b\Delta^2  }{2v} \, e^{-2vt}$ and then solving for $C$ at $t=0$, 
\begin{align} 
  C &=    e^{  \frac{ {b\Delta^2} }{2v} } \,\,\sum_{c=0}^{N-1} X_{c,N},   
\end{align}
gives us the solution to the initial value problem for $x_N(t)$:
\begin{align} 
  x_N(t) &=  \, e^{(G_{T_N} -\delta )t } \times \, e^{\frac{ {b\Delta^2} }{2v} (e^{-2vt} - 1)  } \,\times \sum_{c=0}^{N-1} X_{c,N} - e^{-\delta t} \sum_{c=0}^{N-1} X_{c,N}    
\end{align}

If we wish to know the total population abundance, $x_{\mathrm{total}}$, then we could simply add $\sum_0^{N-1} x_c(t) $ to the expression above for $x_N(t)$. Since each previous cohort is decaying by $e^{-\delta t}$, the total abundance of all previous cohorts at time $t$ is simply $ e^{-\delta t}\sum_0^{N-1}  X_{c,N} $. Thus, the total abundance, $x_{\mathrm{total}} = x_n(t) +  \sum_0^{N-1} x_c(t) $,  will be
\begin{align} \label{eq:x_total_homogeneous}
  x_{\textrm{total}} &=  \, e^{(G_{T_N} -\delta )t } \times \, e^{\frac{ {b\Delta^2} }{2v} (e^{-2vt} - 1)  } \,\times \sum_{c=0}^{N-1} X_{c,N} 
\end{align}

If we further wish to write out the expressions for each of the initial values $X_{c,N}$ explicitly in terms of the parameters associated with previous environmental fluctuations, then we can solve Eq. \eqref{eq:x_total_homogeneous} iteratively, starting with the abundance of the population at the end of the first temperature state, then iteratively work up to the abundance associated with the $N-1$ temperature state:  
\begin{align} 
  x_{\mathrm{\,total}}(t) \,&=\,  X_0 \,\, e^{(G_{T_N} -\delta )t} \times \exp\left\{ \, \frac{b_c\Delta_N^2}{2v} \left( e^{-2vt}  -  1 \right) \right. \notag \\ 
  &\left. \quad\quad\quad\quad\quad + \sum_{c=1}^{N-1} \left[ (G_{T_c} -\delta )\tau_c +  \frac{b_c\Delta_c^2}{2v} \left( e^{-2v\tau_c}  -  1 \right) \right] \right\}, 
\end{align}

\subsection*{Derivation by solving the separable linear differential equation}
Since the whole population in the homogeneous case is composed of identical individuals, the IVP can be written out as 
\begin{align} 
  \frac{ d x_{\mathrm{total}}}{dt} \,&=\,   g_N \cdot x_{\mathrm{total}}(t) - \delta \cdot x_{\mathrm{total}}(t), \\
  &  x_{\mathrm{total}}(0)= X_{\mathrm{total}, N},
\end{align}
where $x_{\mathrm{total}}(t)$ is the total abundance of the current (homogeneous) population. This ODE is now a simple linear separable equation 
\begin{align} 
  \frac{ d x_{\mathrm{total}}}{dt} \,&=\,   (G_{T_N} -b (T_N-T_A^\circ)^2 e^{-2vt}) \, x_{\mathrm{total}}(t) - \delta \cdot x_{\mathrm{total}}(t) \notag \\
  &=\,   (G_{T_N}  - b \Delta_N^2 e^{-2vt} -\delta ) \, x_{\mathrm{total}}(t),  \notag 
\end{align}
which can be solved as
\begin{align} 
   \int \frac{ 1}{x} dx \,&=\,  \int (G_{T_N} - b (T_N-T_A^\circ)^2 e^{-2vt} - \delta) \, dt \notag \\
  \Longrightarrow  \quad x_{\mathrm{total}}(t)  &= \exp\left\{ (G_{T_N} - \delta)t + \frac{ b\Delta_N^2  }{2v} \, e^{-2vt} + C \right\}.
\end{align}
After solving for $C$  (where $C = \ln X_{\mathrm{total}, N} - \frac{ b\Delta_N^2  }{2v}$) we get the following expression for $x_{\mathrm{total}}(t)$: 
\begin{align} 
  x_{\mathrm{total}}(t)  &= X_{_{\mathrm{total}, N} }\,\, e^{(G_{T_N} - \delta)t} \times \exp\left\{  \frac{ b\Delta_N^2  }{2v} ( e^{-2vt} -1)\right\},
\end{align}
which is identical to Eq. \eqref{eq:x_total_homogeneous} above. As before, we can write out the expression for $X_{\mathrm{total}, N}$ explicitly in terms of the  parameters associated with previous environmental states by solving the above equation for $x_{\mathrm{total}}(t)$ iteratively, either backwards (starting with the $N-1$ environmental state), or forwards (starting with the first environmental state).

\section*{Appendix C: Parameters for case studies}
The time series of the population densities plotted in Figure 2 were obtained by simulating a series of 20 temperature fluctuations using a square wave function, where temperatures oscillated between $14^{\circ}$C and $30^{\circ}$C. The time series started with a population that was assumed to be acclimated at $T_0 = 14^{\circ}$C before subjecting it to an initial temperature of $T_1 = 30^{\circ}$C. The parameters used in plotting the time series in Figure 2 are given below in Table C1.
\bigskip
\begin{center}
\begin{tabular}{lc}
  
  \toprule
  \multicolumn{2}{c}{\textbf{Table C1: Parameter values for Fig. 2}} \\
  \midrule
  Parameter         & Value \\
  \midrule
  $N$                                    & 20 \\
  $\delta$                               & 0.35  \\
  $v$                                    & 1.3  \\
  $\tau_k$                               & 1 day\\
  $X_{0,1}$                              & 2 \\
  $T_{\mathrm{min}}$                     & $14^{\circ}$C \\
  $T_{\mathrm{max}}$                     & $30^{\circ}$C \\
  $G_{T_N}$ (when $T_N=14^{\circ}$C)     & 0.88\\
  $G_{T_N}$ (when $T_N=30^{\circ}$C)     & 2.34\\
  \bottomrule
\end{tabular}
\end{center}

The value for the parameter constant $b_i$ (where $i$ is the $i^{th}$ fluctuation or state) depends on  whether $i$ represents an odd or even temperature state in the time series: 
\begin{align}
  b_{i}  =  
  \begin{cases}  \frac{G_{14^{\circ}\text{C}} - 0.38}{ (T_{\mathrm{max}} -T_{\mathrm{min}})^2 }& \text{when $i$ is odd}, \\           
    \frac{G_{30^{\circ}\text{C}} - 2.64}{ (T_{\mathrm{max}} -T_{\mathrm{min}})^2 } &\text{when $i$ is even}. 
  \end{cases}  
\end{align}
The $b_i$ expression used for plotting the time series in Figure 2 is similar to a function used by Kremer et al.'s \citep{kremer.fey.ea_18}. The values for $G_{T_N}$ when $T_N= 14^{\circ} $C and $30^{\circ}$C (last two entries in Table C1), and the numerical values 0.38 and 2.64 found in the expression for $b_i$ above, were obtained from Kremer et al.'s Appendix B, Table B1.

\end{document}